\documentclass[twocolumn,showpacs,amsfonts,aps,prc,floatfix,%
superscriptaddress,nofootinbib]{revtex4}

\usepackage{amsmath}
\usepackage{bm}
\usepackage{graphicx}
\usepackage{url}
\usepackage{longtable}

\newcommand{\be}[1]{\begin{equation}\label{#1}}
\newcommand{\ee}{\end{equation}}

\def\d{d}
\def\D{D}

\begin{document}


\title{
Relativistic fluctuating hydrodynamics
with memory functions and colored noises
}
\date{\today}

\author{Koichi Murase}
\email{murase@nt.phys.s.u-tokyo.ac.jp}
\affiliation{Department of Physics, the University of Tokyo,
Tokyo 113-0033, Japan}
\affiliation{Theoretical Research Division, Nishina Center, RIKEN,
Wako 351-0198, Japan}
\affiliation{Department of Physics, Sophia University,
Tokyo 102-8554, Japan}

\author{Tetsufumi Hirano}
\email{hirano@sophia.ac.jp}
\affiliation{Department of Physics, Sophia University,
Tokyo 102-8554, Japan}

\begin{abstract}
Relativistic dissipative hydrodynamics including
hydrodynamic fluctuations is formulated
by putting an emphasis on non-linearity and causality.
As a consequence of causality, 
dissipative currents become
dynamical variables
and
noises appeared in an integral form of
constitutive equations
should be colored ones from fluctuation-dissipation
relations.
Nevertheless noises turn out to be white ones
in its differential form when
noises are assumed to be Gaussian.
The obtained differential equations
are very useful in numerical implementation
of relativistic fluctuating hydrodynamics.
\end{abstract}

\maketitle



\textit{Introduction}---
Relativistic hydrodynamics has been widely used so far
in various fields such as cosmology, astro- and nuclear physics
to describe space-time evolution of
thermodynamic variables and flow phenomena.
In the physics of relativistic heavy ion collisions,
relativistic hydrodynamic models play
a crucial role in drawing a remarkable
conclusion that the quark gluon plasma (QGP) \cite{Yagi:2005yb},
which is novel deconfined nuclear matter
and filled the early universe,
is created very quickly after the collisions \cite{Heinz:2001xi}, 
records the highest man-made
temperature of the order of $10^{12}$K \cite{Adare:2008ab,guiness},
and, most interestingly, behaves like a perfect fluid \cite{BNL}.
A new paradigm of strongly coupled
QGP was established through 
a vast body of efforts 
of both the experimental 
\cite{Arsene:2004fa,Back:2004je,Adams:2005dq,Adcox:2004mh}
and theoretical analyses \cite{sQGP1,sQGP2,sQGP3,Hirano:2005wx}.
The next task in this field is to precisely extract transport coefficients
of the QGP such as shear and bulk viscosities
through comparison of results from relativistic hydrodynamic
simulations of the QGP
with experimental data intimately related with collective flow phenomena.

In ordinary hydrodynamic framework,
thermodynamics is assumed to be applicable in a local rest frame
at each space-time point.
This implicitly means that
there exists an intermediate scale at which 
the size of a fluid element can be characterized.
At this scale, a microscopic characteristic scale 
should be small enough to apply thermodynamics in a fluid element,
while macroscopic characteristic
scale
should be much larger than that scale.
In order to obtain a set of equations of motion and constitutive
equations,
one can introduce the Knudsen number, $K=l/L$,
where $l$ is the microscopic scale and $L$
is the macroscopic scale,
and expand macroscopic quantities
in a series of the small parameter $K$.
In this way,
hydrodynamics has been applied to
macroscopic system to describe slow and long-wave-length-limit dynamics.
Hydrodynamics is applicable when
separation of these three scales can be justified.

Since thermodynamics can be derived
by taking an ensemble average of many 
canonical systems in statistical physics,
this concept lies implicitly also in hydrodynamics.
In the physics of relativistic heavy ion collisions,
hydrodynamic simulations 
had been performed
from smooth initial fields
\cite{Huovinen:2003fa,Kolb:2003dz}. 
The resultant dynamics is regarded as an average behavior of many similar events
with almost the same impact parameter.
Recently, event-by-event hydrodynamic simulations 
have been performed extensively to describe experimentally measured
momentum anisotropy of hadrons \cite{Hirano:2012kj}.
When the QGP is created $\sim1$ fm/$c$ just after the collisions,
causality does not allow the coarse-grained scale to be larger than $\sim 1$ fm.
Consequently the system responds hydrodynamically to the granular structure
originating from fluctuating configuration of nucleons inside colliding nuclei.
However, it is far from trivial whether thermodynamic and hydrodynamic concepts
within its current form
can be compatible with event-by-event description of relativistic heavy ion collisions.
Since information of the system
is reduced in the coarse-grained process,
thermodynamic variables
fluctuate from event to event around average values,
which is nothing but thermal fluctuation.
Therefore the thermal fluctuation
should be taken into account towards a consistent description of
dynamics of thermodynamic fields on an event-by-event basis.

The thermal fluctuations which arise in hydrodynamic evolution
independently for each spatial and temporal points
are called hydrodynamic fluctuations.
A non-relativistic theory of hydrodynamic fluctuation
was initiated by Landau and Lifshitz long time ago \cite{LL,LP}.
They extended the Navier-Stokes 
equations by implementing stochastic fluxes into constitutive equations
based on fluctuation-dissipation relations.
The theory treats \textit{linear} fluctuations around equilibrium state.
There have been several attempts to study
\textit{non-linear} hydrodynamic fluctuation within
the framework of Fokker-Planck equation \cite{ZubarevMorozov}
or that of Langevin equation \cite{Espanol}.
Non-linearities are of particular importance for fluids undergoing, \textit{e.g.},
phase transition, nucleation, and instabilities since
these can amplify the effects of fluctuation exponentially \cite{LBNL1,LBNL2}.

The effect of thermal fluctuation
on transport coefficients 
in the relativistic system was discussed in Refs.~\cite{Kovtun:2011np,PeraltaRamos:2011es}.
An extension of linear hydrodynamic fluctuation in the Landau-Lifshitz theory \cite{LL,LP} 
to the relativistic hydrodynamics
was discussed in Refs.~\cite{Calzetta:1997aj,Kapusta:2011gt,Kumar:2013twa}.
Hydrodynamic fluctuations
on top of
one-dimensionally expanding background field
together with its implication to the physics of relativistic heavy ion collisions
were also investigated \cite{Kapusta:2011gt,Kumar:2013twa}.

On the other hand, a naive 
relativistic extension of the Navier-Stokes equation
\cite{Eckart:1940te, LL}
has a problem on causality \cite{Hiscock:1985zz}.
The second order correction terms to the entropy current
play an essential role for the theory to obey
the causality \cite{Israel:1976tn,Israel:1979wp}.
Thus the non-linear version of relativistic fluctuating hydrodynamics being consistent with causality
 is demanded
for the purpose of event-by-event 
hydrodynamic simulations of, \textit{e.g.}, relativistic heavy ion collisions.
In this Letter, we formulate relativistic fluctuating hydrodynamics 
by putting emphasis on the importance of causality, its consequence in the property of hydrodynamic fluctuation,
and non-linearity of the resulting equations of motion.
In the following, the Minkowski metric is $g^{\mu \nu} = \mathrm{diag}(+, -, -, -)$ and the natural unit $\hbar = c = k_{B} = 1$ is employed.


\textit{Memory function and fluctuation-dissipation relation}---
Hydrodynamic fluctuation is deviation of the dissipative currents,
such as bulk pressure and shear stresses,
from their average values determined by conventional constitutive equations.
Hence the effect of hydrodynamic fluctuation
appears only quadratic (and in general higher order) terms of hydrodynamic fields
in the \textit{linearized} fluctuating hydrodynamics.
However it is far from trivial what happens
when hydrodynamic
\textit{non-linear} equations
 are solved by taking account of hydrodynamic fluctuations.
The hydrodynamic fluctuations are treated as 
stochastic processes
in the hydrodynamic equations. Thus the equations are no longer
deterministic ones, but 
stochastic differential equations
like Langevin equations.
Here we consider the Gaussian noise as the hydrodynamic fluctuations.
The power spectrum of the hydrodynamic fluctuations
is related to transport properties
through the {fluctuation-dissipation relation}.

To determine the power spectrum of the hydrodynamic fluctuations
in relativistic dissipative hydrodynamics with finite relaxation times,
we can write down the constitutive equations
explicitly 
with respect to the dissipative currents
\begin{align}
\Pi &= -\int_{x^0>x'^0} \d^4x' G_\Pi(x-x')
  \theta(x') +\delta\Pi,\\
\pi^{\mu\nu} &= \int_{x^0>x'^0} \d^4x' G_\pi(x-x')^{\mu\nu\alpha\beta}
  (\partial_{\langle\alpha}u_{\beta\rangle}|_{x'}) +\delta\pi^{\mu\nu},\\
\nu_i^\mu &= -\int_{x^0>x'^0} \d^4x' G_{ij}(x-x')^{\mu\alpha}
  (T \nabla_\alpha \textstyle\frac{\mu_j}T|_{x'}) + \delta\nu_i^\mu,
\end{align}
where $u^\mu$ is the four-velocity in the Landau frame \cite{LL},
and $T$ and $\mu_i$ are temperature and chemical potential
of the $i$-th conserved current ($i=1,\ldots,n$), respectively.
The dissipative currents $\Pi$, $\pi^{\mu\nu}$, and $\nu_i^\mu$
are bulk pressure, shear stress tensor,
and diffusion of the $i$-th conserved charge, respectively.
The corresponding thermodynamic forces
are $\theta=\partial_\lambda u^\lambda$,
$\partial_{\langle\alpha}u_{\beta\rangle} = \Delta_{\alpha\beta\gamma\delta}\partial^\gamma u^\delta$,
and $T\nabla_\alpha\tfrac{\mu_j}{T} =T \Delta_{\alpha\beta} \partial^\beta\tfrac{\mu_j}{T}$, 
where $\Delta^{\alpha\beta} =g^{\alpha\beta}-u^\alpha u^\beta$ and
$\Delta^{\mu\nu\alpha\beta}
  = \tfrac12(\Delta^{\mu\alpha}\Delta^{\nu\beta}
  + \Delta^{\mu\beta}\Delta^{\nu\alpha})
  - \frac13\Delta^{\mu\nu}\Delta^{\alpha\beta}$
are projectors to each tensor component of the dissipative currents.
As above, the dissipative currents
can be generally written in the form of
the average behavior of the response from the thermodynamic force
plus its 
hydrodynamic fluctuation
$\delta\Pi$, $\delta\pi^{\mu\nu}$, and $\delta\nu_i^\mu$.
The average behavior can be expressed
as a convolution of the thermodynamic force
and the retarded Green's function.
The Green's function contains the information about
transport properties of the system
such as the shear viscosity $\eta$, the bulk viscosity $\zeta$,
and the charge conductivities $\kappa_{ij}$.
The function also contains the memory effect of the system
such as relaxation time $\tau_{R}$.
Thus it is also called {memory function}.

Using the fluctuation-dissipation theorem,
we can obtain the power spectrum or the
two point correlation of the hydrodynamic fluctuations as
\begin{align}
  \langle\delta\Pi(x) \delta\Pi(x')\rangle
  &= T G^*_\Pi(x-x'),\\
  \langle\delta\pi^{\mu\nu}(x) \delta\pi^{\alpha\beta}(x')\rangle
  &= \textstyle T G^*_\pi(x-x')^{\mu\nu\alpha\beta},\\
  \langle\delta\nu_i^\mu(x) \delta\nu_j^\alpha(x')\rangle
  &= T G^*_{ij}(x-x')^{\mu\alpha}.
\end{align}
Here the memory functions are
extended for the domain of $x'^0<x^0$
to be even functions, {\it e.g.},
$G^*_{ij}(x)^{\mu\alpha} = G^*_{ji}(-x)^{\alpha\mu}
= G_{ij}(x)^{\mu\alpha} + G_{ji}(-x)^{\alpha\mu}$.

In the case of
the first-order dissipative hydrodynamics
with hydrodynamic fluctuation,
$G^*_{\Pi}(x) = 2\zeta \delta^{(4)}(x)$,
$G^*_{\pi}(x)^{\mu\nu\alpha\beta}
  = 4\eta \delta^{(4)}(x) \Delta^{\mu\nu\alpha\beta}$,
and
$G^*_{ij}(x)^{\mu\alpha}
  = -2\kappa_{ij} \delta^{(4)}(x) \Delta^{\mu \alpha}$.
These memory functions are local in time and have no memory effects:
The correlation of the fluctuation
vanishes
in an infinitesimally small time duration.
Thus the fluctuations exhibit \textit{white noises} in the first-order 
dissipative hydrodynamics.

Note that, in a more general case which respects the causality,
the memory function should vanish
with two spatially separate points, {\it i.e.},
$G(x-x')=0$ where $(x-x')^2<0$,
while the delta functions in the first-order case
are symmetric in any direction of $x-x'$.


\textit{A second-order dissipative hydrodynamics with hydrodynamic fluctuations}---
As mentioned before,
the first-order relativistic dissipative hydrodynamics
violates causality \cite{Hiscock:1985zz}. 
To respect the causality in the relativistic theories,
second or higher order dissipative hydrodynamics 
with relaxation effects
should be considered.
Here we consider a simple case
of the second-order constitutive equations \cite{Israel:1976tn,Israel:1979wp}
\begin{align}
  \tau_\Pi \D \Pi +\Pi
    &= -\zeta\theta,
    \label{eq:simple.constitutive.1}\\
  \tau_\pi \Delta^{\mu\nu}{}_{\alpha\beta}\D\pi^{\alpha\beta} +\pi^{\mu\nu}
    &= 2\eta \partial^{\langle\mu} u^{\nu\rangle},
    \label{eq:simple.constitutive.2}\\
  \tau_{ij} \Delta^{\mu}{}_{\alpha}\D \nu_j^\alpha +\nu_i^\mu
    &= \kappa_{ij}T \nabla^\mu \frac{\mu_j}T,
    \label{eq:simple.constitutive.3}
\end{align}
where $\D = u^\alpha \partial_\alpha$ is the time derivative in the Landau frame.
The memory functions can be obtained by solving
the constitutive equations for dissipative currents \cite{Lublinsky:2009kv}
\def\ExponentialGreenFunction#1#2{\frac1{#2}\exp\left({-\frac{#1}{#2}}\right)}%
\def\tFin{\tau_\mathrm{f}}%
\def\tIni{\tau_\mathrm{i}}%
\def\Fin{_\mathrm{f}}%
\def\Ini{_\mathrm{i}}%
\begin{align}
G_{\Pi}(x-x')
  =& \zeta \ExponentialGreenFunction{\tau-\tau'}{\tau_\Pi} \theta^{(4)}(x-x'),
    \label{eq:memory.bulk}\\
G_{\pi}(x-x')^{\mu\nu\alpha\beta}
  =& 2\eta\ExponentialGreenFunction{\tau-\tau'}{\tau_\pi}\nonumber\\
   & \times\Delta(\tau;\tau')^{\mu\nu\alpha\beta}\theta^{(4)}(x-x'),
    \label{eq:memory.shear}\\
G_{ij}(x-x')^{\mu\alpha}
  =& \tau^{-1}_{ij}
     \left[\mathrm{T}\exp\left({-\int_{\tau'}^\tau \d\tau'' \tau^{-1}_{jk}|_{\tau''} }\right)\right]_{jk}
     \kappa_{kl}\nonumber\\
   & \times\Delta(\tau;\tau')^{\mu\alpha}\theta^{(4)}(x-x')
    \label{eq:memory.diff},
\end{align}
where $\sigma^\mu=(\tau(x), \boldsymbol{\sigma}(x))$ are
the proper time and the co-moving coordinates in the Landau frame, respectively,
and $\mathrm{T}\exp({-\int_{\tau'}^\tau \d\tau'' \tau^{-1}_{jk}|_{\tau''} })$
is the time-ordered exponential.
The function $\theta^{(4)}(x-x') = \left|\tfrac{\partial\sigma^\mu}{\partial x^\nu}\right|
\delta^{(3)}(\bm{\sigma}-\bm{\sigma}')\Theta(\tau-\tau')$
is defined as the common part of the memory functions.
The tensors $\Delta(\tFin;\tIni)^{\mu\nu\alpha\beta}$
and $\Delta(\tFin;\tIni)^{\mu\alpha}$
are the time-by-time projection
into the tensor components of the dissipative currents
and defined as follows:
\begin{align}
&\Delta(\tFin;\tIni)^{\mu\nu}{}_{\alpha\beta}
  = \lim_{N\to\infty}
    \Delta(\tFin)^{\mu\nu}{}_{\alpha_0\beta_0}
  \nonumber\\&\quad\times
    \left[\prod_{k=0}^{N-1}
      \Delta(\tFin + \tfrac{\tIni-\tFin}N k)
      ^{\alpha_{k}\beta_{k}}{}_{\alpha_{k+1}\beta_{k+1}}\right]
    \Delta(\tIni)^{\alpha_N\beta_N}{}_{\alpha\beta},
    \label{eq:simple.pi.f.def}\\
&\Delta(\tFin;\tIni)^{\mu}{}_{\alpha}\nonumber
  = \lim_{N\to\infty}
    \Delta(\tFin)^{\mu}{}_{\alpha_0}
  \nonumber\\&\quad\times
   \left[\prod_{k=0}^{N-1}
      \Delta(\tFin + \tfrac{\tIni-\tFin}N k)^{\alpha_{k}}{}_{\alpha_{k+1}} \right]
    \Delta(\tIni)^{\alpha_N}{}_{\alpha}.
\end{align}
This non-trivial tensor structure comes from the
projections $\Delta^{\mu\nu}{}_{\alpha\beta}$ and
$\Delta^{\mu}{}_{\alpha}$ in the first term in the left hand sides of
constitutive equations
(\ref{eq:simple.constitutive.2}) and
(\ref{eq:simple.constitutive.3}).
This time-by-time projection is required
to make $\pi^{\mu\nu}$ and $\nu_i^\mu$ constrained
to each tensor space during time evolution.

In these memory functions (\ref{eq:memory.bulk})-(\ref{eq:memory.diff}),
the time correlation has a form of exponential relaxation.
In general,
the memory function of causal relativistic dissipative hydrodynamics
has non-vanishing values for different times.
Consequently, the time correlation of the fluctuations $TG(x)$
does not vanish between different times, \textit{i.e.},
the fluctuations exhibit \textit{colored noises}.
For the bulk pressure,
colored noises are manifested in the Fourier space as 
  \begin{align}
     \langle\delta\Pi^*_{\omega,\bm{k}}\delta\Pi_{\omega',\bm{k}'}\rangle
        &= \frac{(2\pi)^4 \delta(\omega-\omega') \delta^{(3)}(\bm{k}-\bm{k}')}{(\omega\tau_\Pi)^2 + 1}.
  \end{align}
Here  ($\omega,\,\bm{k})$ are the conjugate variables for $(\tau,\,\bm{\sigma})$,
respectively.
For shear stress tensor and diffusion, the explicit formulas
are much more involved due to time-by-time projection. Nevertheless,
the corresponding power spectra in the Fourier space
depend on $\omega$ similar to the bulk pressure case. 


\textit{Implementation in numerical simulations}---
There are two different strategies to
implement hydrodynamic fluctuations in numerical simulations.
One is the way to memorize the past fluid fields and 
to directly solve the constitutive equations with integral.
The other is to go back
to the differential form of constitutive equation
such as Eqs.~(\ref{eq:simple.constitutive.1})-(\ref{eq:simple.constitutive.3})
with  the hydrodynamic fluctuations.

Since 
constitutive equations
containing memory functions are non-Markovian,
solving the equations requires the information about the all past fluid fields.
So, in numerical simulations, one has to memorize the past fields of the all steps.
Nevertheless one can introduce some cutoff step number of the past field to be memorized
since the memory function relaxes to zero in finite time
and the system should ``forget'' the information about past near the equilibrium.

In the case of the fluctuating hydrodynamics
with second-order dissipative terms discussed above,
the constitutive equations become
\begin{align}
\Pi =& -\zeta \int^{\tau}_{-\infty} \d\tau'
  \ExponentialGreenFunction{\tau-\tau'}{\tau_\Pi}
  \theta(x') +\delta\Pi,\\
\pi^{\mu\nu} 
  =& 2\eta \int^{\tau}_{-\infty} \d\tau'
  \ExponentialGreenFunction{\tau-\tau'}{\tau_\pi}
  \nonumber\\&\times
  \Delta(\tau;\tau')^{\mu\nu\alpha\beta}
  (\partial_{\langle\alpha}u_{\beta\rangle}|_{x'}) +\delta\pi^{\mu\nu},\\
\nu_i^\mu =& -\int^{\tau}_{-\infty} \d\tau'
  \tau^{-1}_{ij} \left[\mathrm{T}\exp\left(-\int_{\tau'}^\tau
     \d\tau'' \tau^{-1}_{jk}|_{\tau''} \right)\right]_{jk} \kappa_{kl}
  \nonumber\\&\times
  \Delta(\tau;\tau')^{\mu\alpha}
  (T \nabla_\alpha \textstyle\frac{\mu_j}T|_{x'}) + \delta\nu_i^\mu.
\end{align}

It is difficult to evaluate exactly
the time-ordered exponential
and time-by-time projections,
$\Delta(\tFin;\tIni)^{\mu\nu\alpha\beta}$
and $\Delta(\tFin;\tIni)^{\mu\alpha}$.
So one can use the approximated expressions
for $\pi^{\mu\nu}$ and $\nu_i^\mu$
\begin{align}
\pi^{\mu\nu} =& 2\eta\Delta^{\mu\nu\alpha\beta} \int^{\tau}_{-\infty} \d\tau'
  \ExponentialGreenFunction{\tau-\tau'}{\tau_\pi}
  \nonumber\\&\times
  (\partial_{\langle\alpha}u_{\beta\rangle}|_{x'}) +\delta\pi^{\mu\nu},\\
\nu_i^\mu =& -\Delta^{\mu\alpha} \int^{\tau}_{-\infty} \d\tau'
  \tau^{-1}_{ij} [\exp(-\tau^{-1}_{jk}|_{\tau} (\tau'-\tau) )]_{jk} \kappa_{kl}
  \nonumber\\&\times
  (T \nabla_\alpha \textstyle\frac{\mu_j}T|_{x'}) + \delta\nu_i^\mu.
\end{align}
These equations are consistent with
the constitutive equations of the differential form,
(\ref{eq:simple.constitutive.1})-(\ref{eq:simple.constitutive.3}),
up to the second order in thermodynamic forces.

A more practical way of performing numerical calculations
is to employ differential forms of the constitutive equations.
For the simplest case of the second-order dissipative hydrodynamics,
the constitutive equations including hydrodynamic fluctuations become
\begin{align}
  \tau_\Pi \D \Pi +\Pi
    &= -\zeta\theta +\xi_\Pi,\\
  \tau_\pi \Delta^{\mu\nu}{}_{\alpha\beta}\D\pi^{\alpha\beta} +\pi^{\mu\nu}
    &= 2\eta \partial^{\langle\mu} u^{\nu\rangle} +\xi^{\mu\nu}_{\pi},\\
  \tau_{ij} \Delta^{\mu}{}_{\alpha}\D \nu_j^\alpha +\nu_i^\mu
    &= \kappa_{ij}T \nabla^\mu \frac{\mu_j}T + \xi^\mu_{i}.
\end{align}
Those equations are solved by regarding
the dissipative currents as the dynamical variables
in addition to the other fluid fields.
This is the consequence of causality and the
essential difference between the first order and
the second order theories.
It is shown that
the noise terms in the constitutive equations 
in the differential form above become white ones
even though the hydrodynamic fluctuation itself is colored one:
\begin{align}
  \langle\xi_{\Pi}(x) \xi_{\Pi}(x')\rangle
  &= 2 T \zeta \delta^{(4)}(x-x'),\\
  \langle\xi_\pi^{\mu\nu}(x) \xi_\pi^{\alpha\beta}(x')\rangle
  &= 4 T \eta \delta^{(4)}(x-x')\Delta^{\mu\nu\alpha\beta},\\
  \langle\xi_i^\mu(x) \xi_j^\alpha(x')\rangle
  &= 2 T \kappa_{ij}\delta^{(4)}(x-x')\Delta^{\mu\alpha}.
\end{align}
Further more, it is also shown 
that, if the hydrodynamic fluctuations are Gaussian noises,
the differential form of constitutive equations should have the following 
general form in the Fourier space:
\begin{equation}
  [i\omega A_{\bm{k}} + 1] \Pi'_{\omega,\bm{k}} = \kappa F_{\omega,\bm{k}} + \xi_{\omega,\bm{k}}.
\end{equation}
Here $\Pi'$ is a dissipative current
and $\kappa F$ is the first-order term of the thermodynamic force in 
the constitutive equation.
Also, the noise term $\xi_{\omega,\bm{k}}\, (= [i\omega A_{\bm{k}} + 1] \delta\Pi'_{\omega,\bm{k}})$
becomes Gaussian white noise:
\begin{equation}
  \langle\xi(x)\xi(x')\rangle = 2 T \kappa \delta^{(4)}(x-x').
\end{equation}
Therefore it is much more convenient to solve the equations in the differential
form in numerical calculations
rather than the ones in its integral form if and only if a general assumption that
noises are Gaussian can be made. 
The proof will be discussed elsewhere
in the future publication \cite{MuraseHirano}.


\textit{Conclusions}---
To discuss about fluctuating hydrodynamics, which is the hydrodynamics
with dissipation and the corresponding hydrodynamic fluctuations,
in relativistic system, one should consider the memory effects
such as relaxation times to respect the causality.
The constitutive equations can be written down in
the form of the response of thermodynamic forces in the present and past,
and the response functions are called memory functions.
This is the generalization of the conventional first-order or second-order
constitutive equations and they are the special case with certain memory functions.
The power spectrum of the hydrodynamic fluctuations is related 
to the memory function by fluctuation-dissipation relation.
It should be noted that the memory function can be obtained
in principle from the first principle calculations such as lattice
quantum chromodynamics \cite{Nakamura:2004sy,Meyer:2007dy}
through correlation function of energy momentum tensor
in equilibrium.
In the case of the first-order dissipative hydrodynamics,
the fluctuations are white noises
while in the case of the second-order dissipative hydrodynamics
respecting causality, the fluctuations should be colored.
If the hydrodynamic fluctuations are Gaussian noises,
the differential form of the constitutive equations 
within the second-order relativistic dissipative hydrodynamics
should have the specific structure,
and the corresponding noises turn out to be white ones.

Instead of conventional relativistic dissipative
hydrodynamic equations,
the equations obtained in this Letter can be employed 
to simulate 
more realistically, \textit{e.g.},
space-time evolution of the QGP 
in relativistic heavy ion collisions on an event-by-event basis.
Numerical implementation of these equations together with
the effects of hydrodynamic fluctuations 
on observables will be discussed elsewhere \cite{MuraseHirano}.


We acknowledge fruitful discussion with A.~Monnai and Y.~Hirono
at the early stage of the present work.
The work of K.M. is supported by the Japan Society for the
Promotion of Science for Young Scientists.
The work of T.H. is partly supported by
Grant-in-Aid for Scientific Research
No.~25400269.


\end{document}